\documentclass[preprint2]{aastex631}
\begin{document}
\title{Ultracool and Brown Dwarf Candidates Identified via CatWISE-2MASS-Gaia DR3}

\author[0000-0001-8803-3840]{Daniel Majaess}
\affiliation{Mount Saint Vincent University, Halifax, Canada}
\email{Daniel.Majaess@msvu.ca}

\author[0000-0002-7064-099X]{Dante Minniti}
\affiliation{Instituto de Astrofísica, Dep.~de Física y Astronomía, Facultad de Ciencias Exactas, Universidad Andres Bello, Av.~Fernández Concha 700, Santiago, Chile}
\affiliation{Vatican Observatory, Specola Vaticana, V-00120, Vatican City, Vatican City State}

\author[0000-0002-4430-9427]{Matias Gomez}
\affiliation{Instituto de Astrofísica, Dep.~de Física y Astronomía, Facultad de Ciencias Exactas, Universidad Andres Bello, Av.~Fernández Concha 700, Santiago, Chile}

\author[0000-0001-6878-8648]{Roberto K. Saito}
\affiliation{Departamento de Física, Universidade Federal de Santa Catarina, Trindade 88040-900, Florianópolis, Brazil}

\author[0000-0002-1860-2304]{Maria G. Navarro}
\affiliation{INAF - Osservatorio Astronomico di Roma, Via di Frascati 33, 00078 Monte Porzio Catone, Italy}

\begin{abstract}
Ultracool and brown dwarf candidates ($d_{\odot}\lesssim 125$ pc, $|b|>8^{\circ}$) were identified via multiband and astrometric data from CatWISE ($W_1W_2$), 2MASS ($JHK_s$), and Gaia DR3 ($G$, $G_{RP}$, $\pi$, $\mu_{\alpha}$, $\mu_{\delta}$). $N\approx 11.8 \times 10^3$ candidates emerged once simultaneously constrained by color and absolute magnitude criteria (e.g., $G-W_2 \ge 1.75(G-J)-2.25$), whereby $\simeq 350$ sources are absent from the Gaia Ultracool Dwarf (UCD) catalog. Spectral types were approximated using a hybrid $M_G-M_{W_1}$ sigmoid that offers additional temperature coverage in certain cases.  Subsequent efforts may focus on extending sampling to the deeper NIR VVVX footprint that partly encompasses the Galactic plane, and over the long-term spectroscopically (in)validating candidates missing from the Gaia UCD database.
\end{abstract}

\section{Introduction}
Ultracool dwarfs occupy the M7V and later spectroscopic regime \citep[e.g.,][]{kir97}.  Continued census efforts may facilitate delineating the hydrogen-deuterium fusion boundary and other insightful transitions \citep[e.g.,][]{die14}, and more broadly, brown dwarfs are lucrative targets for exoplanet search campaigns \citep[e.g.,][]{del18}.  Indeed, the multiplanetary TRAPPIST-1 system features an M8V host \citep[][]{gil17}.  Toward expanding the population of known ultracool dwarfs, for example, \citet{rey18} exploited Gaia DR2 to identify 679 new L-dwarf candidates. \citet{sma19} likewise employed DR2 and detailed 695 targets (M8–T6), of which 20 were newly discovered multiple systems. \citet{gol23} compiled the Fifth Catalogue of Nearby Stars (CNS5), and characterized 701 brown dwarfs within $d<25$ pc.  

\begin{figure}[t]
 \centering
  \includegraphics[width=0.99\columnwidth]{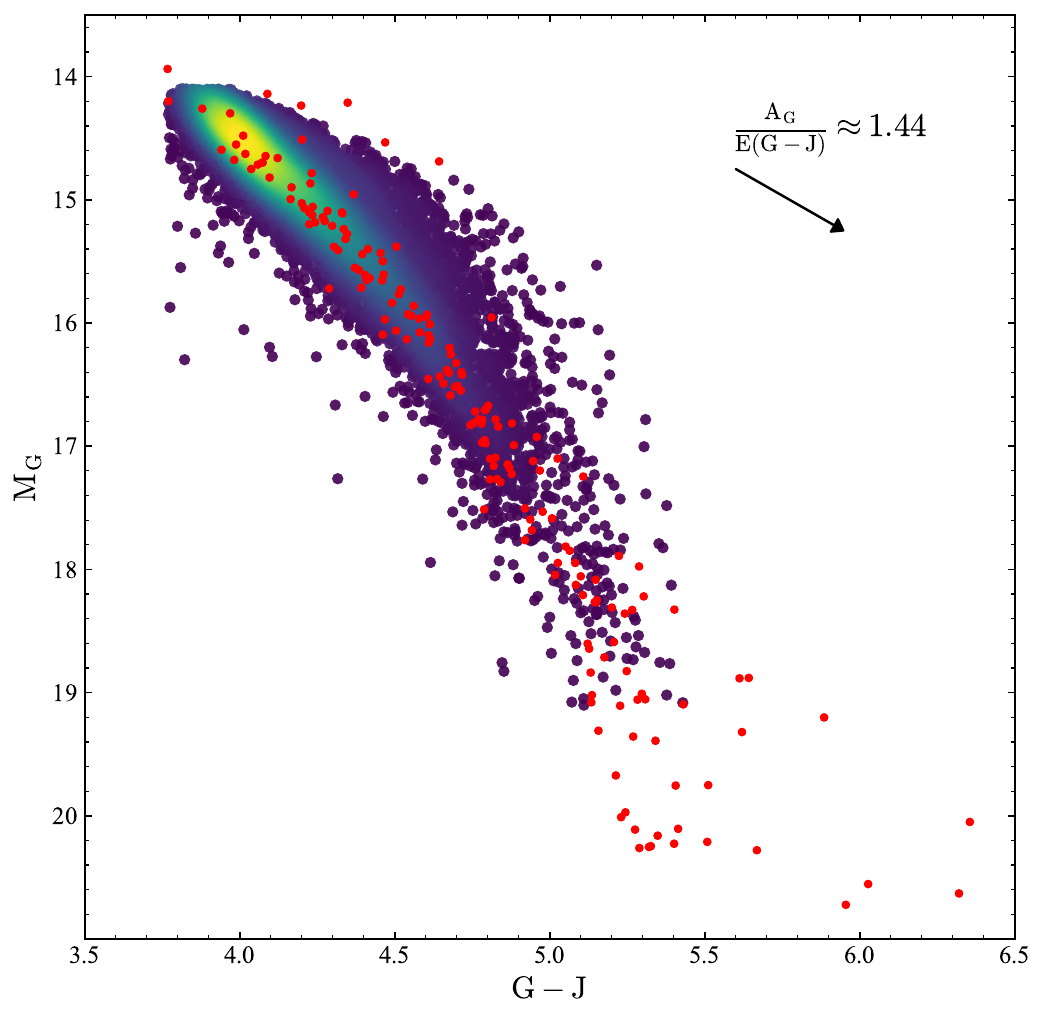} 
  \caption{Ultracool \textit{candidates} identified by pairing MIR CatWISE, NIR 2MASS, and optical Gaia DR3 ($N\approx11.8 \times 10^3$, $d_{\odot}\lesssim 125$ pc, $|b|>8^{\circ}$), and adopting conservative photometric and astrometric criteria to mitigate false-positives (\S \ref{sec-analysis}).  $N\simeq350$ sources are absent from the Gaia UCD database (Fig.~\ref{fig-missing}), while 11402 targets are found therein. The candidate distribution aligns with the trend delineated by the red datapoints \citep[M7 and cooler,][]{kir24}.}
 \label{fig-cmd}
\end{figure}

\begin{figure*}[t]
\centering
  \includegraphics[width=0.99\columnwidth]{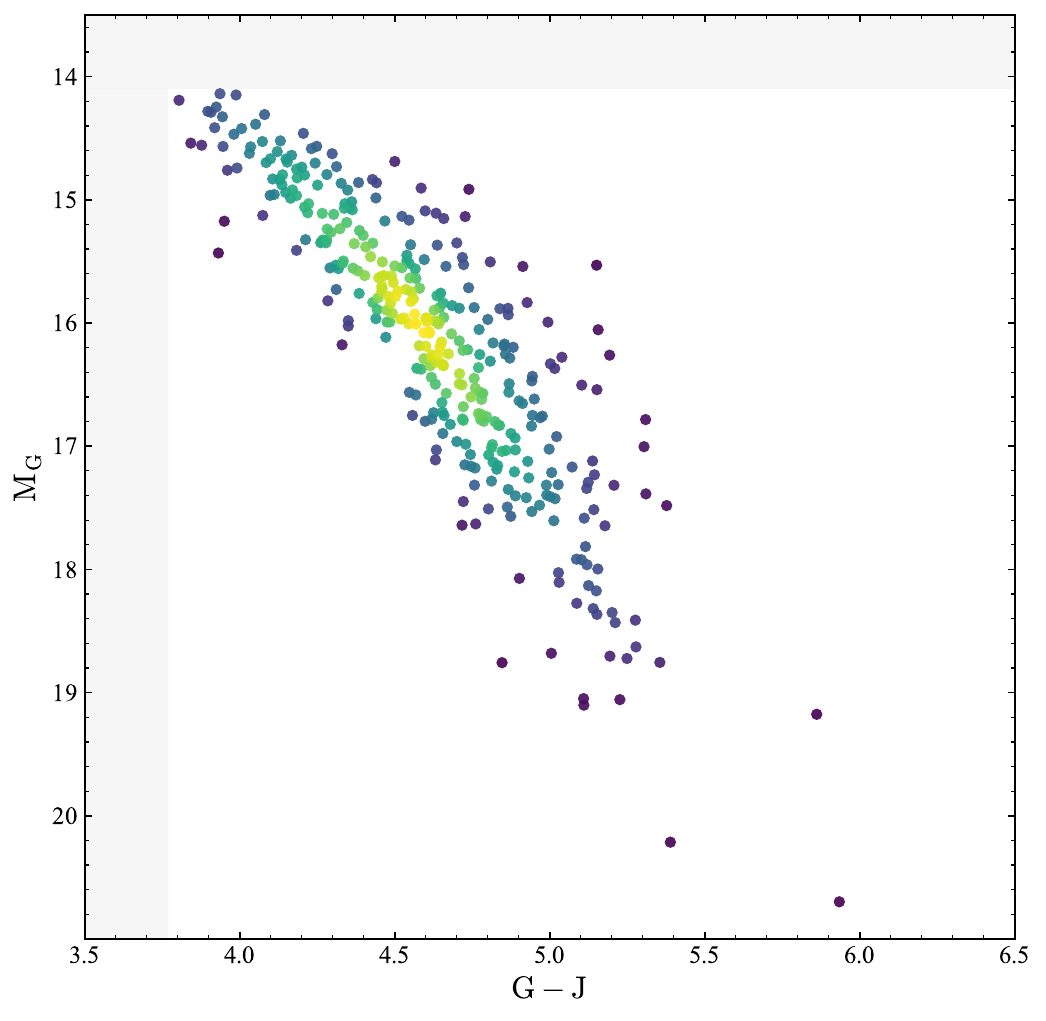} 
  \includegraphics[width=0.99\columnwidth]{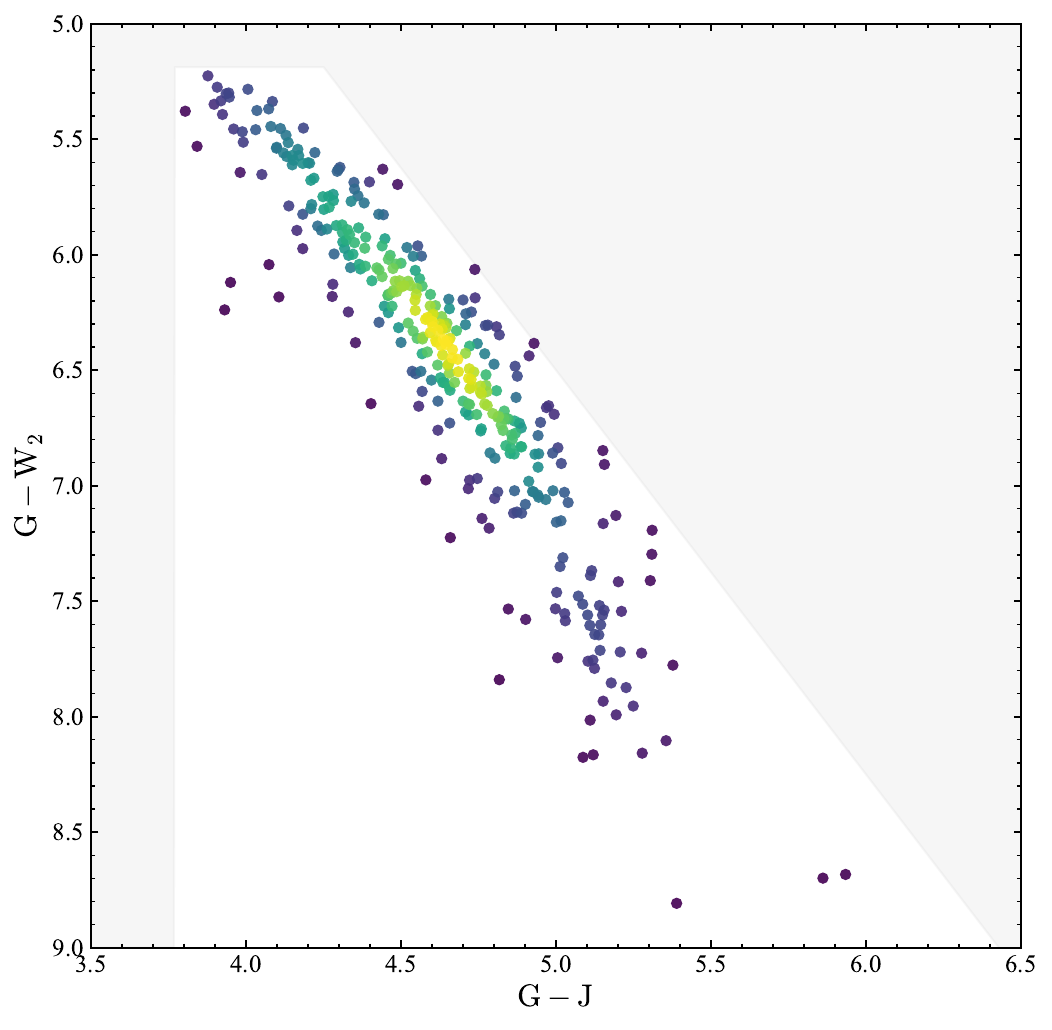}
  \caption{Ultracool dwarf candidates absent from the Gaia UCD.  Possibly $\simeq 77$\% are not featured in complementary catalogs examined, however, there may be research the authors are inevitably unfamiliar with, and duplicates may exist.  Long-term spectroscopic follow-up shall proceed for brighter targets (e.g., a potential offshoot of the envisioned KMOS VVVX-GalCen initiative), which will eliminate false-positives emerging solely from photometry and astrometry, and owing to confusion arising from the sampling radius. Gray regions represent color and absolute magnitude cuts, and the dearth in candidates arises from faintness limits (e.g., 2MASS $J\lesssim 16^{m}.3$, and for Gaia $G$ see Fig.~5 in \citealt{rey24}).  Hence an advantage of VVVX and Euclid.}
 \label{fig-missing}
\end{figure*}

\begin{figure}[t]
\centering
  \includegraphics[width=0.99\columnwidth]{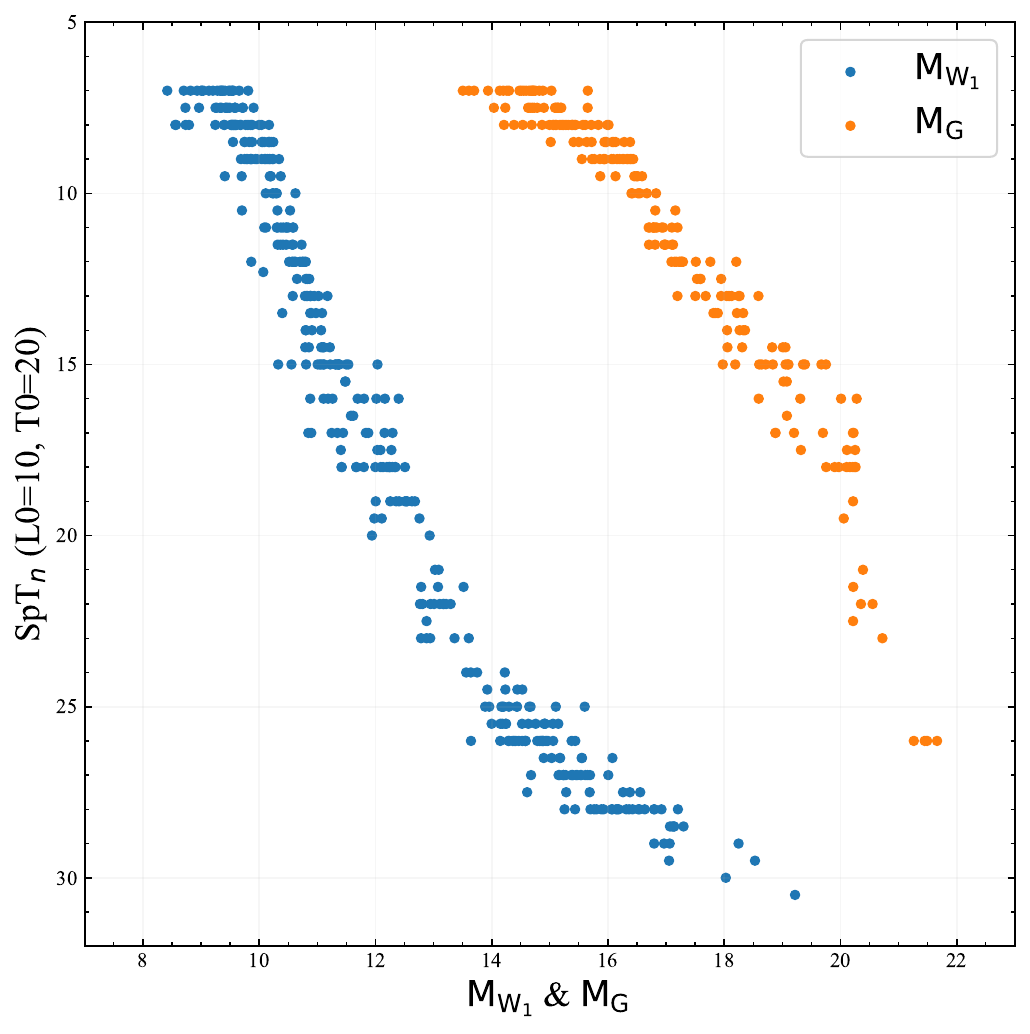}
  \caption{$M_{W_1}$ (blue datapoints) can provide spectral coverage toward cooler brown dwarfs. A hybrid $M_G-M_{W_1}$ function (see text) was constructed to provide \textit{approximate} spectral types for the ultracool dwarf catalog assembled.  A sigmoid regulates the changeover from $M_{G}$ for hotter candidates to $M_{W_1}$ for cooler targets (Eqn.~\ref{sigmoid}). The subsample presented stems from \citet{kir24}.}
 \label{fig-w1}
\end{figure}

In this study, CatWISE ($W_1W_2$, \citealt{mar21}), 2MASS ($JHK_s$, \citealt{cut03}), and Gaia DR3 ($G$, $G_{RP}$, $\pi$, $\mu_{\alpha}$, $\mu_{\delta}$, \citealt{gai23}) are mobilized to expand the population of ultracool dwarfs \citep[e.g., add to the Gaia Ultracool Dwarf (UCD) catalog,][and references therein]{cre23a,sar23}. An emphasis is placed on leveraging multiband and astrometric criteria to mitigate false-positives.  CatWISE provides enhanced $W_1W_2$ photometry, and for example, \citet{mar21} note the temporal baseline is markedly enhanced relative to ALLWISE.  Gaia provides key astrometric constraints which fosters matching to absolute magnitudes endemic to the ultracool domain \citep[e.g.,][DR2]{rey18}.  Specifically, \citet[][$20$ pc sample]{kir24} supplied a calibrating dataset that is exploited to help unveil ultracool dwarfs via photometric cuts (Figs.~\ref{fig-cmd}, \ref{fig-missing}), and is utilized to construct a hybrid spectral type estimator across the temperature domain (Fig.~\ref{fig-w1}).  

\section{Analysis}
\label{sec-analysis}
Photometric culling criteria stemmed from an analysis of the \citet{kir24} dataset (Fig.~\ref{fig-cmd}), and in particular the M7V designated LSR J0515$+$5911. Breadth was afforded for photometric uncertainties and peculiarities by adopting a $0^{\rm m}.2$ buffer, relative to that star's absolute magnitudes ($G$, $G_{RP}$, $J$, $H$, $K_s$, $W_1$, $W_2$). Similarly, the implied colors $G-G_{RP}$, $G-J$, $G-W_1$, $J-K_s$ were likewise adjusted by that amount (e.g., $J-K_s\gtrsim 4.5$). Subsequent cuts of $W_1-W_2>0.1$ and $G-W_2 \ge 1.75(G-J)-2.25$ were implemented to mitigate a redward $G-J$ plume \citep[see also \S 2.1 of][]{sar23}. 2MASS photometry was restricted to QFlg$=$AAA, and uncertainties in all bands were required. The forthcoming Euclid survey shall provide a marked improvement relative to 2MASS limits (e.g., $\Delta J\simeq 7^{m}.5$), although with less spatial coverage (i.e., $\simeq \frac{1}{3}$ all-sky).  Indeed, the aforementioned faintness constraints on T dwarfs are apparent in Fig.~\ref{fig-missing} \citep[see also][their Fig.~5]{rey24}, and a subset of culling criteria are highlighted thereon.

The following constraints were adopted to reduce parallax uncertainties from offsetting absolute magnitudes: $\pi > 8$ mas ($d_{\odot}\lesssim 125$), ${\rm RUWE}< 1.4$, and ${\rm RPlx}> 8$ (\textit{parallax over error}, e.g., $10 \sigma$ was selected by \citealt{rey18}). The RUWE criterion ensured that the optimal 5-parameter Gaia solution was utilized, and hence all candidates feature proper motions. A positional criterion of $|b|>8^{\circ}$ was imposed to lessen confusion from different objects, and minimize the introduction of hotter reddened stars.  A separate initiative will build upon existing efforts \citep{bea14,mej22} to investigate ultracool dwarfs within the VVV/VVVX footprint \citep[i.e., covering portions of the Galactic plane,][]{min11,sai12,sai24}.  For example, \citet{mej22} indicated the following could be ultracool dwarfs: Gaia EDR3 5697747496563028352 (M9), 5990656908867751808 (M7), 6062219761378864128 (M7), 5257810972787599360 (M7), 5409811755160786048 (M7), 5323745691499188736 (M7), 5693793442288873984 (M7).

A $25^{\prime\prime}$ matching radius was selected between Gaia and 2MASS, and thereafter relative to CatWISE. The criterion bolstered the detection of higher proper motion dwarfs. Sources that passed the aforementioned suite of criteria were classified as ultracool candidates (Fig.~\ref{fig-cmd}). Reassuringly, the curvature of the ultracool sequence emerges, and since it is unveiled primarily by limits inferred from an upper-end M7V calibrator:~that implies the sample is  primarily \textit{bona fide} ultracool dwarfs. Indeed, 71\% overlap with the Gaia UCD catalog's golden subsample \citep{cre23b}, 11402 from a total of 11753 sources presented are in the broader UCD database (i.e., 97\%, whereby the full UCD catalog extends to $\pi>1.7$ mas and features $94\times 10^3$ entries).  That points to $\simeq 350$ missing datapoints.  Relative to \citet{rey18}, \citet{sma19}, and \citet[][CNS5]{gol23}, from the missing sources 42, 27, and 12 overlap, accordingly. $N\simeq276$ remain after removing duplicates, and the mode is approximately a magnitude fainter than the overlapping UCD sample (i.e., $G\simeq20^{m}.6$). That subset partly emerges from DR3 improvements and additional sampling of fainter targets, relative to DR2. 

Fig.~\ref{fig-cmd} hosts $\approx 11.8 \times 10^3$ candidates, and the data will be made available as a CDS VizieR catalog. Table~\ref{table} relays a subset of the information to be presented, such as a column indicating whether a target is featured in the Gaia UCD golden sample (UCD\_golden), the broader Gaia UCD (UCD), \citet[][R18]{rey18}, \citet[][S19]{sma19}, \citet[][CNS5]{gol23}, or potentially missing ($-$). Admittedly, there is research the authors are invariably unaware of, and overlap may exist.  $T_{\rm eff} \pm \Delta T_{\rm eff}$ from the Gaia UCD will be added to the VizieR catalog for overlapping entries, and rounded to the nearest integer.

Spectral type approximations were deduced via a sigmoid weighted $M_G-M_{W_1}$ function, in tandem with estimates from \citet{rey18} for overlapping entries. $M_{W_1}$ provides complimentary constraints owing to increasing coverage across later spectral types (blue datapoints in Fig.~\ref{fig-w1}). The sigmoid issues hotter dwarfs spectral types inferred from $M_G$, and thereafter a smooth transition occurs across early-L ($M_{W_1}=10.3$) where estimates are from both $M_G$ and $M_{W_1}$, and subsequently cooler candidates rely on $M_{W_1}$: 
\begin{eqnarray}
\label{sigmoid}
{\rm SpT}_n &=&w(42.52 - 5.70 M_G + 0.225 M_G^2)+(1-w) \nonumber \\
     &&\times(-56.94 + 9.02 M_{W_1} - 0.233 M_{W_1}^2)    \\
    w &=&(1 + e^{10(M_{W_1} - 10.3)})^{-1} \nonumber
\end{eqnarray}
M7V was adopted as the spectral type index number ${\rm SpT}_n=7$, L0 is 10, and T0 is 20 \citep[see also][]{rey18}. The cited spectral type is merely indicative (Table~\ref{table}), and the approach may be unsuitable for other objectives.

A reddening vector $A_G/E(G-J)\approx1.44$ is displayed in Fig.~\ref{fig-cmd} \citep[][their Table 3]{wc19}.  Importantly, an effort was made to minimize the impact of dust obscuration by instituting the $|b|>8^{\circ}$ cutoff, together with an upper distance limit ($d<125$ pc, and the median is $\simeq 70$ pc). Confusion arising from a background source in concert with reddening may foster false-positives for a comparatively marginal subsample, an assertion underpinned by the dataset naturally adhering to the intrinsic curvature of the calibrating sample (Figs.~\ref{fig-cmd},~\ref{fig-missing}).

\begin{deluxetable*}{lccccccccccc}
\tablecaption{Subsample of ultracool dwarfs.\label{table}}
\tablehead{
\colhead{Gaia DR3} & \colhead{$\pi \atop (mas)$} & \colhead{${\mu \atop (mas\ yr^{-1})}$} & \colhead{G} & \colhead{$G_{RP}$} & \colhead{J} & \colhead{H} & \colhead{K$_s$} & \colhead{W$_1$} & \colhead{W$_2$} & \colhead{Ref.} & \colhead{SpT}}
\startdata
4703216410071320832 & 17.6 & 104 & 20.67 & 19.14 & 16.02 & 15.15 & 14.57 & 14.37 & 14.10 & R18 & L2 \\
2308116769195247360 & 12.4 & 186 & 20.49 & 18.98 & 15.96 & 15.29 & 14.76 & 14.61 & 14.36 & $-$ & M9 \\
2423078853137423488 & 23.0 & 179 & 20.55 & 18.95 & 15.68 & 14.82 & 14.50 & 13.88 & 13.52 & $-$ & L3 \\
5178144483388552832 & 13.0 & 344 & 20.61 & 19.07 & 16.03 & 15.42 & 14.85 & 14.59 & 14.33 & $-$ & L0 \\
4722209515662513920 & 18.9 & 458 & 20.66 & 19.05 & 15.81 & 15.12 & 14.68 & 14.17 & 13.95 & $-$ & L2 \\
5107303911284277504 & 32.3 & 135 & 20.63 & 19.04 & 15.48 & 14.44 & 13.86 & 13.34 & 13.07 & R18 & L4 \\
495947643267031040 & 15.2 & 32 & 20.44 & 18.89 & 15.78 & 15.03 & 14.51 & 14.08 & 13.71 & $-$ & L0 \\
284894801473847296 & 11.3 & 31 & 20.61 & 19.10 & 15.97 & 15.29 & 14.88 & 14.52 & 14.30 & $-$ & M9 \\
944440390446800768 & 18.0 & 114 & 20.55 & 19.03 & 15.72 & 14.91 & 14.26 & 13.90 & 13.73 & $-$ & L0 \\
954562326987474304 & 10.8 & 110 & 20.38 & 18.92 & 15.86 & 15.19 & 14.88 & 14.48 & 14.27 & $-$ & M8 \\
3361948739720043776 & 19.4 & 137 & 20.54 & 18.99 & 15.81 & 14.92 & 14.56 & 14.17 & 13.97 & R18 & L3 \\
5193549977029793792 & 14.3 & 53 & 20.42 & 18.85 & 15.82 & 15.17 & 14.45 & 14.37 & 14.13 & $-$ & M9 \\
901999138535923840 & 16.7 & 48 & 20.67 & 19.15 & 15.95 & 15.09 & 14.54 & 14.19 & 13.98 & $-$ & L1 \\
907818338184084096 & 17.6 & 154 & 20.57 & 19.02 & 15.75 & 14.98 & 14.28 & 14.02 & 13.84 & $-$ & L1 \\
587601798685457792 & 20.3 & 167 & 20.53 & 18.92 & 15.79 & 14.89 & 14.34 & 14.11 & 13.84 & $-$ & L3 \\
5443893385806317440 & 18.8 & 123 & 20.13 & 18.57 & 15.41 & 14.56 & 14.30 & 13.65 & 13.48 & R18 & L0 \\
723985800815301248 & 17.3 & 93 & 20.57 & 19.04 & 15.77 & 15.11 & 14.63 & 14.17 & 13.88 & $-$ & L1 \\
3564538223902705280 & 27.1 & 134 & 20.76 & 19.19 & 15.65 & 14.60 & 14.02 & 13.50 & 13.20 & $-$ & L3 \\
3808497893938381824 & 14.7 & 229 & 20.42 & 18.89 & 15.78 & 15.17 & 14.60 & 14.36 & 14.12 & $-$ & L0 \\
3760469851647361664 & 15.6 & 64 & 20.53 & 19.00 & 15.90 & 15.13 & 14.64 & 14.52 & 14.26 & $-$ & L2 \\
3814316135810315904 & 17.1 & 162 & 20.40 & 18.87 & 15.53 & 14.85 & 14.49 & 14.10 & 13.92 & S19 & L0 \\
3812678894277121920 & 29.6 & 291 & 20.77 & 19.15 & 15.65 & 14.71 & 13.95 & 13.41 & 13.13 & $-$ & L3 \\
3917930950515862272 & 17.9 & 267 & 20.47 & 18.93 & 15.70 & 14.93 & 14.42 & 14.05 & 13.82 & $-$ & L1 \\
5379915930922278400 & 11.4 & 34 & 20.68 & 19.19 & 16.08 & 15.30 & 14.78 & 14.56 & 14.37 & $-$ & M9 \\
3893817767165654400 & 29.0 & 242 & 20.68 & 19.11 & 15.53 & 14.55 & 14.02 & 13.45 & 13.14 & S19 & L3 \\
1371557673804322176 & 13.8 & 64 & 20.83 & 19.29 & 16.07 & 15.38 & 14.87 & 14.56 & 14.22 & $-$ & L0 \\
1631078603058453632 & 19.6 & 185 & 20.86 & 19.31 & 16.10 & 15.24 & 14.68 & 14.35 & 14.09 & $-$ & L3 \\
1434750111746029056 & 12.4 & 100 & 20.05 & 18.50 & 15.50 & 14.90 & 14.45 & 14.12 & 13.98 & $-$ & M8 \\
1421716951881370752 & 16.4 & 114 & 20.71 & 19.18 & 15.99 & 15.20 & 14.66 & 14.29 & 14.06 & $-$ & L1 \\
2257578507499901312 & 28.3 & 178 & 20.55 & 18.93 & 15.44 & 14.60 & 14.13 & 13.34 & 13.19 & $-$ & L3 \\
6733294760528587008 & 16.3 & 182 & 19.50 & 17.91 & 14.93 & 14.28 & 14.02 & 13.69 & 13.49 & $-$ & M8 \\
4088155897620238336 & 14.7 & 169 & 20.13 & 18.58 & 15.61 & 14.97 & 14.52 & 14.32 & 14.16 & $-$ & M9 \\
6773385703237821312 & 15.0 & 113 & 20.41 & 18.82 & 15.82 & 14.99 & 14.61 & 14.39 & 14.19 & $-$ & L0 \\
1816087549148430848 & 37.5 & 393 & 20.45 & 18.90 & 15.31 & 14.36 & 13.75 & 13.11 & 12.93 & $-$ & L4 \\
4228935888971277824 & 11.5 & 112 & 20.71 & 19.25 & 16.07 & 15.32 & 14.79 & 14.35 & 14.16 & $-$ & M9 \\
2663586537698797184 & 12.3 & 104 & 20.33 & 18.72 & 15.70 & 14.93 & 14.47 & 14.15 & 13.97 & S19 & M9 \\
2233222229000888064 & 39.7 & 209 & 20.63 & 19.10 & 15.35 & 14.39 & 13.65 & 12.91 & 12.47 & CNS5 & L4 \\
2420682536264210944 & 15.8 & 257 & 20.61 & 18.99 & 15.87 & 15.11 & 14.67 & 14.41 & 14.23 & $-$ & L1 \\
... & &  &  & &  &  &  &  &  &  &  \\
\enddata
\tablenotetext{ }{ Notes: complete findings shall on CDS, and a portion of the database is relayed in terms of quantity and columns (e.g., the VizieR catalog will exhibit astrometric and photometric uncertainties).   References include \citet[][R18]{rey18}, \citet[][S19]{sma19}, and \citet[][CNS5]{gol23}. Photometry stems from 2MASS \citep[][]{cut03}, CatWISE \citep[][]{mar21}, and Gaia DR3 \citep[][]{gai23}.}
\end{deluxetable*}

\section{Conclusions}
A catalog of $N\approx 11.8 \times 10^3$ ultracool and brown dwarf candidates was produced (Figs.~\ref{fig-cmd},~\ref{fig-missing}), and is tied to observations from CatWISE, 2MASS, and Gaia DR3.  \citet{kir24} ultracool dwarf observations were utilized to infer the applied absolute magnitude and photometric color cuts.  This initiative semi-independently contributes to overarching efforts aiming to characterize the relatively local ultracool demographic \citep[e.g.,][]{rey18,sma19,rey21,sar23,kir24,rav24}.  The catalog shall be posted to VizieR (Table~\ref{table}), and contains approximate spectral types inferred from a hybrid $M_G-M_{W_1}$ function (Fig.~\ref{fig-w1}, Eqn.~\ref{sigmoid}). In this instance a sigmoid was implemented to modulate the transition between $M_G$ and $M_{W_1}$ estimators across early-L.

A key next step is beginning a longer-term spectroscopic initiative to (in)validate candidates as done previously \citep[e.g., VVV BD001 and WISE 2121-6239,][]{bea13,bea15}. That undertaking may emerge as an offshoot from the forthcoming KMOS VVVX-GalCen project, and the observations may focus on brighter candidates absent from the Gaia UCD. For example, \citet{bea13} acquired spectra for the L-dwarf VVV BD001 using the Folded-port InfraRed Echellette at the Magellan Baade telescope. Furthermore, bypassing or complementing the optical Gaia G-band faint limit by relying on NIR VVVX ($\pi$ \& $\mu$) could result in expanded ultracool dwarf statistics (e.g., \citealt{bea13} exploited VVV photometry to establish $\pi$ for VVV BD001), and determinations of such parallaxes continue improving. That may partly foster the identification of simultaneous color-color and absolute magnitude-color slope discontinuities across numerous passbands, which could aid demarcate pertinent transitions such as increasing atmospheric condensates, or the hydrogen-deuterium fusion boundary. More broadly atmospheric parameters, metallically dependencies, and binaries may be characterized \citep[e.g.,][]{rav24}. 

However, the deeper NIR VVVX footprint partly encompasses the Galactic plane where false-positives increase from confusion and reddening, and thus enhanced judiciousness is warranted. For example, earlier M-dwarfs near the M7V ultracool boundary may seep into the domain via reddening. For coverage beyond the Galactic plane the Euclid survey shall considerably surpass 2MASS $JH$ limiting magnitudes, and can be paired with Gaia.  Furthermore, the next release of low-resolution Gaia XP spectra shall provide a pertinent dataset for validating candidates \citep[see also][and discussion therein]{sar23}.

\begin{acknowledgments}
\textbf{Acknowledgments}: this research relied on initiatives such as CDS, NASA ADS, arXiv, 2MASS, WISE (CatWISE), Gaia, and staff at the Gaia Help Desk. TapVizieR and Gaia UCD access were facilitated by python and vibcoding in Gemini 3. D.~Minniti gratefully acknowledges support from the Center for Astrophysics and Associated Technologies CATA by the ANID BASAL projects ACE210002 and FB210003, and by Fondecyt Project No.~1220724. R.~K.~S.~acknowledges support from CNPq/Brazil through projects 308298/2022-5 and 421034/2023-8.
\end{acknowledgments}

\bibliography{article}{}
\bibliographystyle{aasjournal}
\end{document}